# Multiband thermal transport in the iron-based superconductor $Ba_{1-x}K_xFe_2As_2$

Marcin Matusiak[1,*] and Thomas Wolf[2]

1. Institute of Low Temperature and Structure Research, Polish Academy of Sciences,

ul. Okolna 2, 50-422 Wroclaw, Poland

2. Institute of Solid State Physics (IFP), Karlsruhe Institute of Technology, D-76021,

Karlsruhe, Germany

**PACS**

74.70.Xa, 74.25.fc, 72.15.Jf

**Abstract**

We present results of precise measurements of the thermal and electrical transport in the optimally- and over-doped $Ba_{1-x}K_xFe_2As_2$ single crystals ($x$ = 0.35, 0.55, 0.88) and compare them to the previously reported data on $Ba(Fe_{1-y}Co_y)_2As_2$. A contraction of the electron pocket is observed upon substitution potassium for barium, but even at the extreme doping ($x$ = 0.88) there is still a noticeable contribution from negative charge carriers to the electronic transport. The size of the electron pocket in all K-doped samples is small enough to cause a significant enhancement of the respective Hall-Lorenz number.

Another observed characteristic is the emergence of a maximum in the transverse thermal conductivity below the superconducting critical temperature of the optimally- ($x$ = 0.35) and slightly over-doped ($x$ = 0.55) samples. The evolution of this anomaly from the optimally electron-doped $Ba(Fe_{0.94}Co_{0.06})_2As_2$ to hole-overdoped $Ba_{0.45}K_{0.55}Fe_2As_2$ suggests formation of a uniform superconducting gap on the electron pocket in the former and regions of a depressed gap on the hole-pocket in the latter.

**Introduction**

There are common characteristics of unconventional superconductors that might suggest a universal physics lying behind the phenomena. The phase diagrams of these compounds generally look alike, often with a magnetically ordered non-superconducting parent compound and a superconducting dome developing upon doping or pressure [1]. The iron-based superconductors are no exception here, but there are some aspects that distinguish them from another numerous family of high-$T_c$ superconductors – copper-based compounds. For example, the ground state of the parent compound in the iron-based materials is weakly metallic [2] instead of being a Mott insulator [3], the Fermi surface is made of hole- and electron-like sheets [4,5] instead of being dominated by one band [6-9], and the pairing symmetry is likely $s_{\pm}$ [10] instead of $d_{x^2-y^2}$ [11]. However, the symmetry of the order parameter in the iron-based superconductors is still under debate, since some reports suggest presence of nodes in the superconducting gap [12] and it is not obvious whether these nodes are either accidental or imposed by the symmetry. Experiments that allow one to investigate the electronic properties in the superconducting state are measurements of the heat transport [13,14], since the thermal signal, unlike the electrical one, is not shorted out by the superconducting condensate. On the other hand, the total thermal conductivity is basically a sum of phonon and electronic contributions and it is not obvious how to separate them. This can be done by performing measurements in a magnetic field [15], when movement of the electrically neutral quasiparticles as phonons remain unaffected, whereas charged ones are subject to the Lorentz force.

In the present work we study the electronic transport properties of single crystals of electron- and hole-doped $BaFe_2As_2$ in the normal and superconducting states. We focus on the relation between the electrical and thermal transport and draw conclusions about the role played by electron- and hole-like pockets in the electronic transport. We also investigate the thermal Hall conductivity in the superconducting phase and see that a significant maximum emerges in K-doped samples. The evolution of the anomaly with doping is interpreted as a result of changes of quasiparticles population in the gapped state.

**Experiment**

$Ba_{1-x}K_xFe_2As_2$ single crystals were grown by a self-flux technique in $Al_2O_3$ crucibles sealed in iron cylinders. The iron cylinders were heated to 950 - 1150°C and subsequently cooled down using very slow cooling rates of 0.3 – 0.6 °C/h. At the end of the growth process the crucibles were tilted in order to decant the remaining flux. The exact K content *x* of the crystals typically differs from the starting composition, and was determined by energy-dispersive x-ray analysis (EDX) and by 4-circle x-ray diffraction.

All transport coefficients were measured along the *ab*-plane of a single crystal with a magnetic field ($B$) applied parallel to the *c* axis. The electrical resistivity ($\rho$) was determined using a four-point technique. The Hall coefficient ($R_H$) measurements were performed by a standard method, where current and magnetic field directions were reversed several times to exclude any influence of the asymmetric position of the Hall contacts and detrimental electromotive forces. The longitudinal ($\kappa_{xx}$) and transverse ($\kappa_{xy}$) thermal conductivities were measured in a single experiment, where respective temperature gradients ($\nabla_x T$ and $\nabla_y T$) were determined with 25 μm Chromel-Constantan (Type E) thermocouples that have been calibrated in a set of magnetic fields. Typically $\nabla_x T$ and $\nabla_y T$ were of the order 1 K/mm and 10 mK/mm, respectively. Measurements were repeated in various magnetic fields between -12.5 T and +12.5 T in order to separate field-symmetric and -antisymmetric components of the signal.

**Results and discussion**

There is a clear difference in the behavior of the electrical resistivity ($\rho$) between the hole- and electron-doped BaFe$_2$As$_2$, as seen in Fig. 1. Not only is the critical temperature ($T_c$ is defined as an inflection point of $\rho(T)$) of the optimally doped Ba$_{0.65}$K$_{0.35}$Fe$_2$As$_2$ (K35), about 14 K higher than $T_c$ of the optimally doped Ba(Fe$_{0.94}$Co$_{0.06}$)$_2$As$_2$ (Co6), but also the shapes of the $\rho(T)$ curves for K- and Co-doped are very different. The increasing potassium content in Ba$_{1-x}$K$_x$Fe$_2$As$_2$ has little effect on the room temperature resistivity, whereas $T_c$ decreases from $T_c$ = 38.8 K in K35, through $T_c$ = 36.8 K in Ba$_{0.45}$K$_{0.55}$Fe$_2$As$_2$ (K55), to $T_c$ = 7.3 K in Ba$_{0.12}$K$_{0.88}$Fe$_2$As$_2$ (K88). Thanks to the suppression of the superconductivity in K88 ($T_c$ < 1.4 K) by a magnetic field of 12.5 T, one can determine the residual resistivity ratio (RRR), which is about 72 in K88, while the RRR ~ 2.5 in Ba(Fe$_{0.76}$Co$_{0.24}$)$_2$As$_2$ (Co24). Such a significant difference is unlikely a consequence of a dramatic change in quality of K- and Co-doped single crystals, since the resistive superconducting transitions remain sharp and similar values of RRR were reported by other groups [16,17]. This rather suggests that electron-like quasiparticles, which are major charge carriers in Ba(Fe$_{1-y}$Co$_y$)$_2$As$_2$, interact more effectively with scattering centers than hole-like ones that dominate electrical transport in Ba$_{1-x}$K$_x$Fe$_2$As$_2$. This is probably because Co substitutes Fe in the conducting FeAs layers while K occupies out-of-plane Ba sites. Studies of the Hall effect in Ba$_{1-x}$K$_x$Fe$_2$As$_2$ by Ohgushi et al. [18] also indicate that heavier holes experience less scattering than electrons, whereas the latter seem to have higher mobility due to a smaller effective mass. As shown in Fig. 2, the Hall coefficient ($R_H$) is negative in the entire temperature range for Ba(Fe$_{1-y}$Co$_y$)$_2$As$_2$ and positive for Ba$_{1-x}$K$_x$Fe$_2$As$_2$ as one can expect from the valence of the dopants. Surprisingly, the room temperature value of $R_H$ in Ba$_{1-x}$K$_x$Fe$_2$As$_2$ does not change much, despite significant changes in the nominal hole concentration, the potassium content. On the other hand, the temperature variation of the Hall coefficient evolves and $R_H(T)$ becomes flatter for higher $x$. These changes were interpreted as a sign of the Lifshitz transition occurring around the critical doping $x = 0.80$ [19], which consists in disappearance of the electron pocket at the M point of the Brillouin zone and the formation of four small hole-lobes. In fact, results of Angle-Resolved Photoemission Spectroscopy (ARPES) measurements confirm that the change of Fermi surface without symmetry breaking takes place between $Ba_{0.6}K_{0.4}Fe_2As_2$ and $KFe_2As_2$ [20]. Other ARPES studies narrow the critical region to $x$: 0.7 – 0.9 [21] or 0.8 – 0.9 [22], though the authors of the latter work do not rule out the possibility that even in $Ba_{0.1}K_{0.9}Fe_2As_2$ one electron band is still crossing the Fermi level and forming a tiny electron pocket at M.

Now we turn to the thermal conductivity ($\kappa_{xx}$) presented in Fig. 3, which is another transport coefficient that is considerably different for $Ba(Fe_{1-y}Co_y)_2As_2$ and $Ba_{1-x}K_xFe_2As_2$. This difference consists not only in the lower thermal conductivity in the normal state of $Ba_{1-x}K_xFe_2As_2$, but also in the behavior of $\kappa_{xx}$ in the superconducting state (right panel in Fig. 3). Namely, the maximum below $T_c$ in $\kappa_{xx}(T)$ of K35 and K55 (where it is virtually identical), is much higher than the one in Co6. Furthermore, in Co6 the maximum is almost completely suppressed by a magnetic field of 12.5 T, despite $T_c$ being 19.5 K. In principle, the maximum below $T_c$ can be of electronic [23-25] or phonon [26,27] origin and it is not trivial to distinguish between these two scenarios. The electronic and phonon contributions can be separated by measuring the transverse (Hall) thermal conductivity ($\kappa_{xy}$) that is also called the thermal Hall or Righi-Leduc effect. The quantity is defined as $\kappa_{xy} / B = \kappa_{xx} \nabla_y T / \nabla_x T$, where $\nabla_y T$ and $\nabla_x T$ are the transverse and longitudinal thermal gradients, respectively. In metals $\kappa_{xy}$ is solely related to the electronic thermal conductivity and includes the sign of the charge carriers. This unique property allows one to use measurements of the transverse thermal conductivity to get an insight into properties of the quasiparticles even deep in the superconducting state [28].

Figure 4 shows that in K35 and K55 a maximum emerges in $\kappa_{xy}(T)$ below $T_c$, which is analogous to the one observed in $\kappa_{xx}(T)$. The presence of such an anomaly in the transverse thermal conductivity was previously reported in the copper-based superconductor $YBa_2Cu_3O_{7-d}$ [29-31] as well as for the heavy fermion $CeCoIn_5$ [32,33]. In both cases it was attributed to an increase in the electronic component of the thermal conductivity arising from lower electron-electron scattering associated with the lower density of states in the superconducting state. As stated by Checkeslky et al. [34] a sizeable thermal Hall signal below the critical temperature is also a sign of a large population of quasiparticles existing in the superconducting state. The authors point to the hole band at the M point of the Brillouin zone (in a 5-band picture) as one with the weakest pairing and the main "source" of quasiparticles below $T_c$. Since a small gap is expected to mimic the behavior of a nodal gap, the hole pocket is suggested to give rise to the anomaly. For a higher potassium content the hole-like signal is predicted to become more pronounced, as a result of possible formation of nodal d-wave order on the hole pockets [35]. As seen in Fig. 4, further K-doping in fact causes a growth of the maximum in $\kappa_{xy}(T)$ in K55 when compared to K35, even if the size of the anomaly in $\kappa_{xx}(T)$ remains the same. On the other hand, cobalt doping of $BaFe_2As_2$ should eventually cause small gap regions on the electron pockets to contribute most of low-energy charge carriers, thus leading to a sign change of the anomaly in the Hall thermal conductivity below $T_c$. However, we find the negative anomaly in Co6 much less pronounced which hints at rather uniform superconducting gap on the electron pocket.

The high temperature part of the thermal Hall signal in all potassium doped samples is very small, i.e. $\kappa_{xy}$ equals approximately $10^{-4}$ W m$^{-1}$ K$^{-1}$, which means that the measured $\Delta_y T$ signal is in the order of a few hundredths of mK per 1 T. The values of $\kappa_{xy}$ in the normal state of K35 and K55 are very similar and undoubtedly negative above $T \sim 140$ K for K35 and $T \sim 160$ K for K55. One may consider this surprising as the Hall coefficients of these samples stay positive in the entire temperature range. In the two-band approximation the total $\kappa_{xy}$ is a sum of the electron-like and hole-like thermal Hall conductivities ($\kappa_{xy}^e$ and $\kappa_{xy}^h$), hence the transverse conductivities ($\sigma_{xy}^e$ and $\sigma_{xy}^h$) weighted by the respective Hall Lorenz numbers ($L_{xy}$):

$$\kappa_{xy} = \kappa_{xy}^e + \kappa_{xy}^h = (L_{xy}^e \sigma_{xy}^e + L_{xy}^h \sigma_{xy}^h)\, T\, (e/k_B)^2, \tag{1}$$

where $L_{xy}$ is defined [15] as

$$L_{xy} \equiv \frac{\kappa_{xy}}{\sigma_{xy} T}\left(\frac{e}{k_B}\right)^2. \tag{2}$$

These lead to

$$L_{xy} = \frac{L_{xy}^e \sigma_{xy}^e + L_{xy}^h \sigma_{xy}^h}{\sigma_{xy}^e + \sigma_{xy}^h}, \tag{3}$$

which means that although for each band $\kappa_{xy}^e/\sigma_{xy}^e$ and $\kappa_{xy}^h/\sigma_{xy}^h$ ratios have to be positive, total $L_{xy}$ may become negative, if $\sigma_{xy}^e$ and $\sigma_{xy}^h$ are comparable but $L_{xy}^e$ is much bigger than $L_{xy}^h$. Figure 5 shows that this is the case of K35 and K55, where the Hall Lorenz numbers are notably different from $L_{xy}(T)$ determined for the Co-doped samples. The electronic transport of the latter seems to be dominated by electron band [36] and the low-temperature maxima in under- and optimally-doped samples were interpreted as a result of the opening of a pseudogap [37]. On the contrary, $L_{xy}(T)$ in K-doped samples are clearly governed by multiband effects.

In Boltzmann transport theory, the transverse transport coefficient are given by [38]:

$$\sigma_{xy} = -2\frac{e^3 B}{\hbar c}\sum_{k}\tau_k v_k^x \left[v_k^y \partial_{k_x} + v_k^x \partial_{k_y}\right]\tau_k v_k^y\left(-\frac{\partial f_k}{\partial \varepsilon_k}\right), \tag{4}$$

and

$$\kappa_{xy} = -2\frac{eB}{T\hbar c}\sum_{k}\varepsilon_k^2 \tau_k v_k^x \left[v_k^y \partial_{k_x} + v_k^x \partial_{k_y}\right]\tau_k v_k^y\left(-\frac{\partial f_k}{\partial \varepsilon_k}\right), \tag{5}$$

where $\varepsilon_k$ is the energy of the quasiparticles, $\tau_k$ is the relaxation time, $v_k$ is the group velocity of the quasiparticles, and $f_k$ is the Fermi distribution function. Their ratio expressed in Eqn. 2 is a "transverse" analog of the Wiedemann-Franz law and, correspondingly, $L_{xy}$ is expected to be equal to $L_0 = \pi^2/3$ for the Fermi liquid in low temperature limit [39]. Retaining a constant relation between the electronic thermal and electrical (multiplied by $T$) conductivities is one of the most fundamental characteristics obeyed by the Fermi liquid. Figure 6 presents the temperature dependences of the Hall Lorenz number calculated numerically for small parabolic pockets of different size. The relaxation time is here assumed to be independent of quasi-momentum. Interestingly, while all $L_{xy}(T)$ curves approach the value of $\pi^2/3$ at $T \to 0$ K as expected, for smaller Fermi energies $L_{xy}$ rises above $\pi^2/3$ and saturates at high temperature at a value larger than $2L_0$. A reason for such a behavior is a large disturbance of the Fermi-Dirac statistics in a conductive band at $T$ higher than the Fermi temperature. A migration of quasiparticles to higher energies and their deficiency below the Fermi level alters the relation between $\kappa_{xy}$ and $\sigma_{xy}$ (as well as $\kappa_{xx}$ and $\sigma_{xx}$) and relatively rises $\kappa_{xy}$ over $\sigma_{xy}T$, hence causes $L_{xy}$ to rise above $L_0$. We need to notice that our numerical result is in some way surprising, since the Fermi-Dirac distribution in the high temperature limit approaches the Maxwell-Boltzmann distribution, and in such a classical case one can expect the value of $L$ to be smaller than $L_0$ [39]. Nevertheless, our data confirm a disproportionately large role of the electron pocket in the thermal transport, which is most likely related to an enhanced value of $L_{xy}^e$. The solid lines presented in Fig. 5 are obtained within the described above two band model (Eqn. 3) under a simple condition that $L_{xy}^h = L_0 = \pi^2/3$ and $L_{xy}^e$ is calculated for a tiny parabolic electron pocket with the Fermi wave vector $k_F = 3.5\%$ $\pi/a$ ($a$ is the lattice constant) for K35, K55 and much smaller $k_F = 2\%$ $\pi/a$ for K88. The number of electrons in the pocket is assumed to be constant, whereas setting the chemical potential fixed would result in steeper $L_{xy}(T)$ for the same value of $k_F$. For the sake of simplicity the $\sigma_{xy}^e/\sigma_{xy}^h$ ratio is assumed to be constant in the entire temperature range. The experimental data for all samples are fitted with the same model, where the Fermi wave vector, as well as the $\sigma_{xy}^e/\sigma_{xy}^h$ ratio were two fitting parameters. These are very simple assumptions, but the main goal was not to match perfectly the experimental curves, but to point at a mechanisms that leads to the atypical electronic transport properties of BaFe$_2$As$_2$, when doped with potassium. Nonetheless, the agreement between the calculated lines and experimental data is satisfactory for K35, K55 and very good for K88. The deviation from the calculated $L_{xy}(T)$ dependence in the low temperature region of K35 and K55 could be related to inelastic scattering in the hole band which suppresses $L_{xy}^h$ or the pseudogap opening that additionally increases $L_{xy}^e$ [37]. The deviation is absent in K88 and one may find surprising that a role of electron pocket in thermal transport is still significant, even it is supposed to be very close to extinction. Our results indicate then that the supposed Lifshitz transition takes place closer to $x = 1$ than thought before and if one takes into consideration the mentioned ARPES reports [21,22] the change in the band structure has to occur at $x$ very close to 0.9.

**Summary**

In summary, we have studied the evolution of the electronic transport properties of potassium-doped Ba$_{1-x}$K$_x$Fe$_2$As$_2$ single crystals from the optimal K35 ($x = 0.35$), through slightly overdoped K55 ($x = 0.55$) to the heavily overdoped K88 ($x = 0.88$) state. A comparison of the results presented here to respective data for Ba(Fe$_{1-y}$Co$_y$)$_2$As$_2$ indicates that while the electronic transport in the latter is dominated by negatively charged carriers, multiband effects are of great importance in Ba$_{1-x}$K$_x$Fe$_2$As$_2$. We also conclude that despite electron-like quasiparticles having higher mobility than hole-like ones, they are subject to stronger scattering. Despite Ba$_{1-x}$K$_x$Fe$_2$As$_2$ being a hole-doped compound an influence from the electron pocket to the thermal transport is evident in all three samples. Namely, we observe a change of sign of the transverse thermal conductivity to negative in K35 and K55, even if the Hall coefficient in these samples remains positive in the entire temperature range. On the other hand, although $\kappa_{xy}$ in K88 stays positive, the temperature dependence of the Hall

Lorenz number in this sample is still significantly affected. The findings are interpreted within the two-band model based on the relaxation time approximation, which is solved numerically. The observed phenomena can be understand as a consequence of the disproportion between values of the Lorenz number for the hole and the electron bands, since our calculations indicate strong enhancement of $L_{xy}$ on a pocket with a very low Fermi level. The presence of the contribution from the electron pocket to the heat transport at the high doping ($x = 0.88$) narrows a possible range of composition where the Lifshitz transition can be anticipated.

Another interesting observation concerns the transverse thermal conductivity anomaly in the superconducting state. Its size grows with the potassium content, which can be a manifestation of the gap nodes emerging on the hole pocket. On the other hand, the very modest minimum in $\kappa_{xy}(T)$ of Ba(Fe$_{1-y}$Co$_y$)$_2$As$_2$ hints at a uniform superconducting gap on the electron pocket.

**Acknowledgments**

The authors would like to thank J.R. Cooper, G. Grissonnanche and L. Spalek for helpful comments.

This work was supported financially by the National Science Centre (Poland) under the research Grant No. 2011/03/B/ST3/00477.

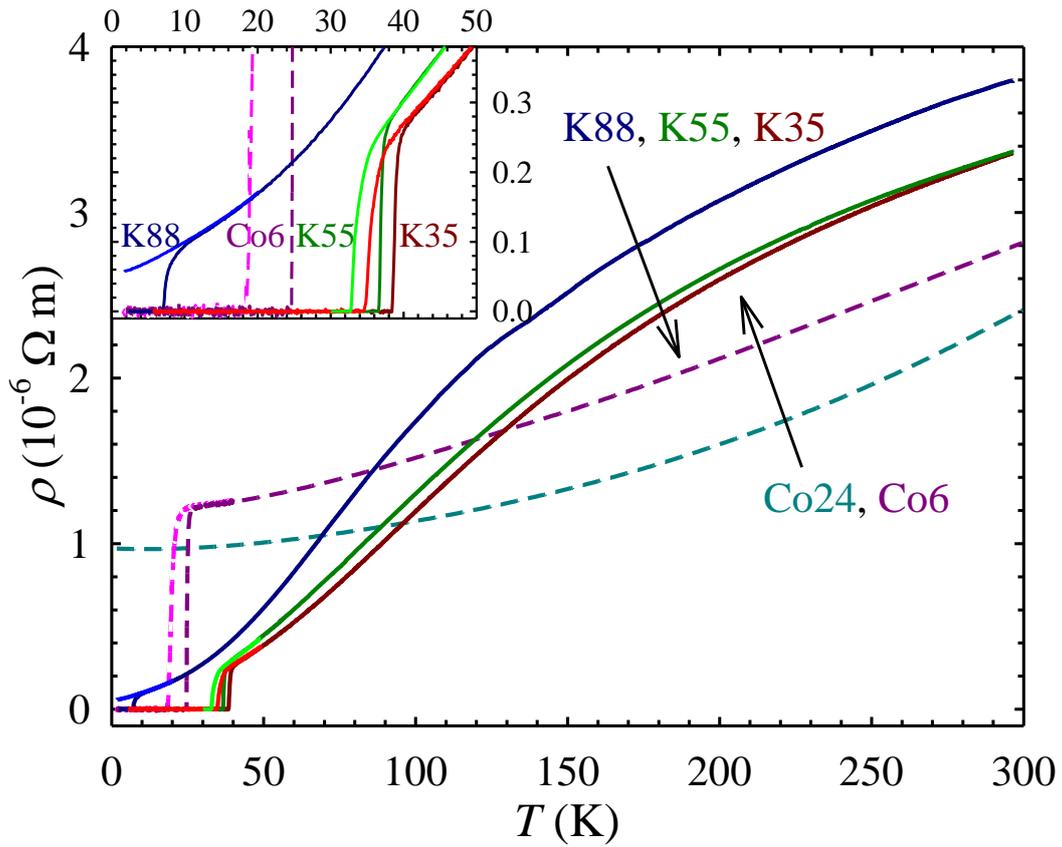

**Figure 1.**

(Color online) The temperature dependences of the resistivity for the $Ba_{1-x}K_xFe_2As_2$ (solid lines) and $Ba(Fe_{1-y}Co_y)_2As_2$ (dashed line) series. Dark lines denote data measured in $B = 0$ T, whereas lighter ones those measured in $B = 12.5$ T. Inset presents the same set of data in the low temperature region.

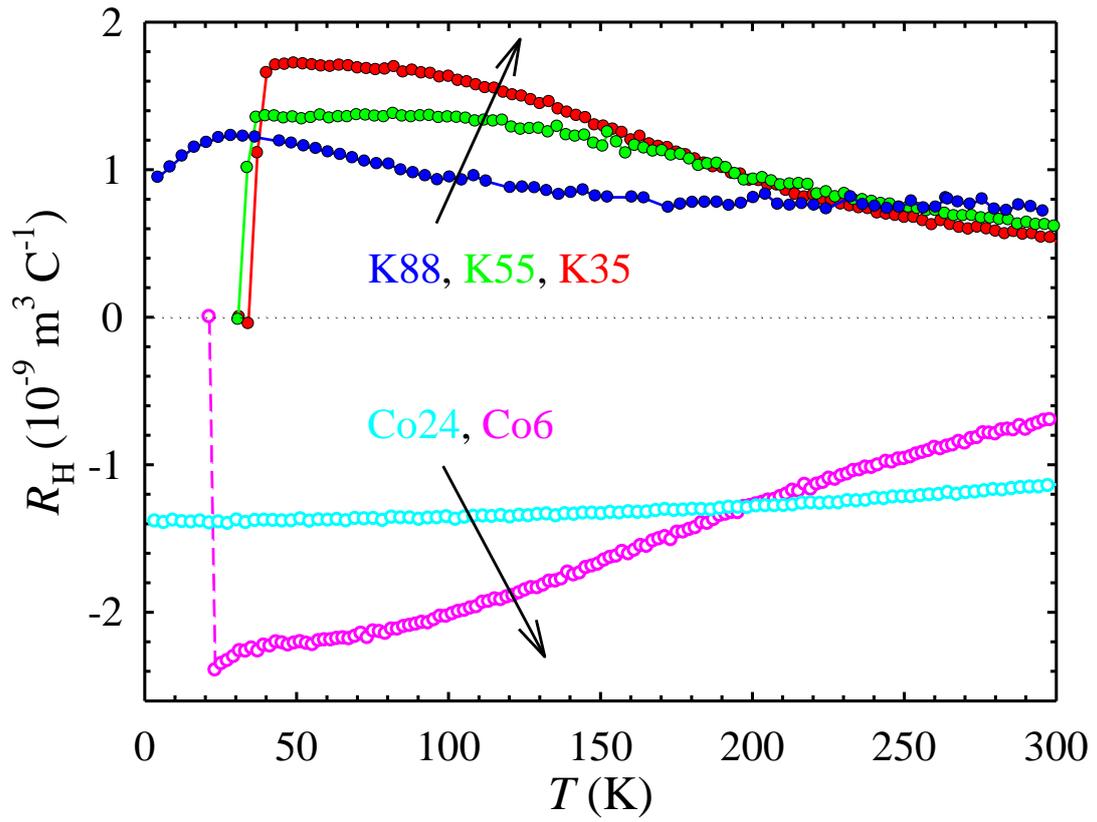

**Figure 2.**

(Color online) The temperature dependences of the Hall coefficient for the $Ba_{1-x}K_xFe_2As_2$ (solid points) and $Ba(Fe_{1-y}Co_y)_2As_2$ (hollow points) series.

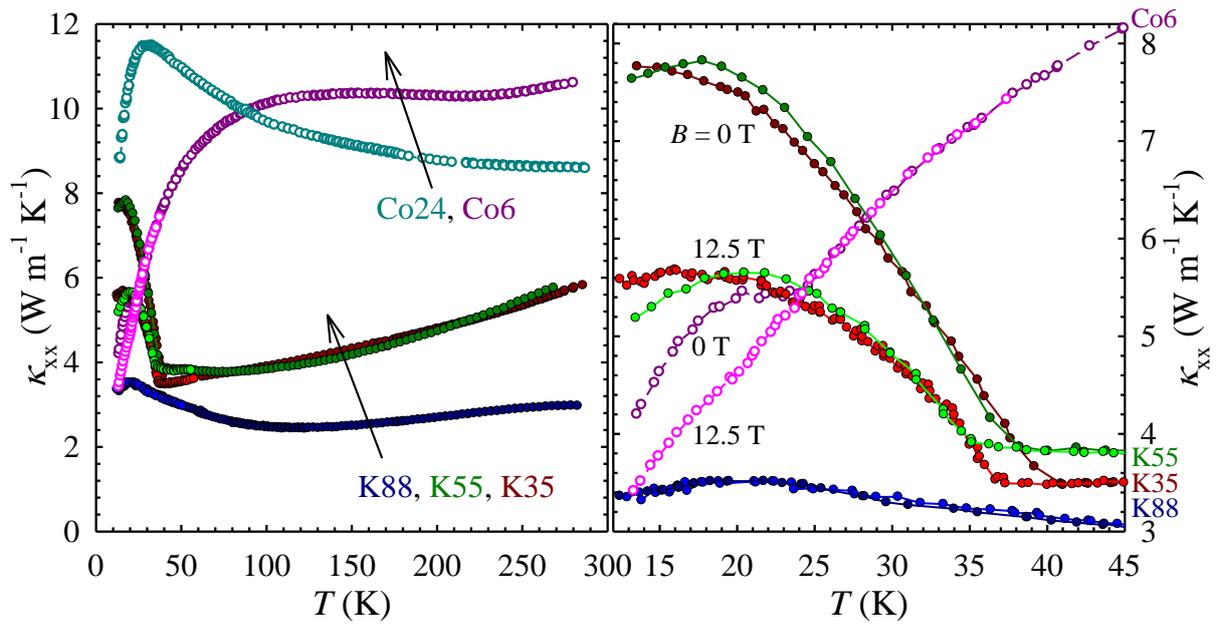

**Figure 3.**

(Color online) Left panel: the temperature dependences of the thermal conductivity for the Ba$_{1-x}$K$_x$Fe$_2$As$_2$ (solid points) and Ba(Fe$_{1-y}$Co$_y$)$_2$As$_2$ (hollow points) series. Dark points denote data measured in $B$ = 0 T, whereas lighter ones those measured in $B$ = 12.5 T. Right panel presents the same set of data in the low temperature region.

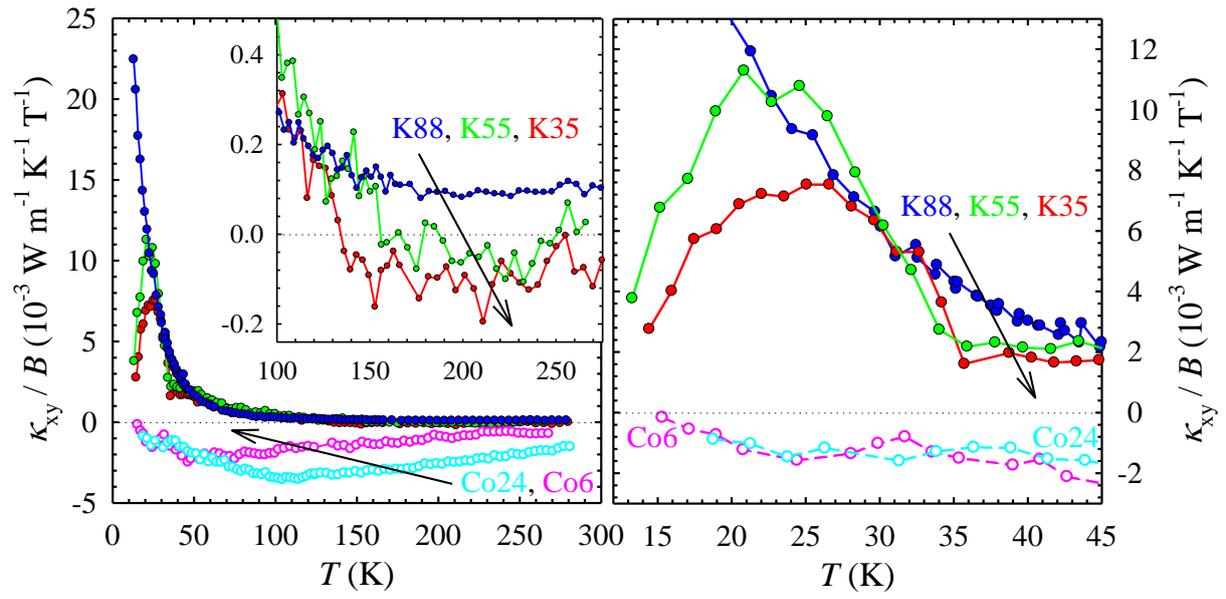

**Figure 4.**

(Color online) Left panel: the temperature dependences of the transverse thermal conductivity measured in $B = 12.5$ T for the $Ba_{1-x}K_xFe_2As_2$ (solid points) and $Ba(Fe_{1-y}Co_y)_2As_2$ (hollow points) series. Inset shows the same set of data in the high temperature region, while the right panel shows the low temperature region.

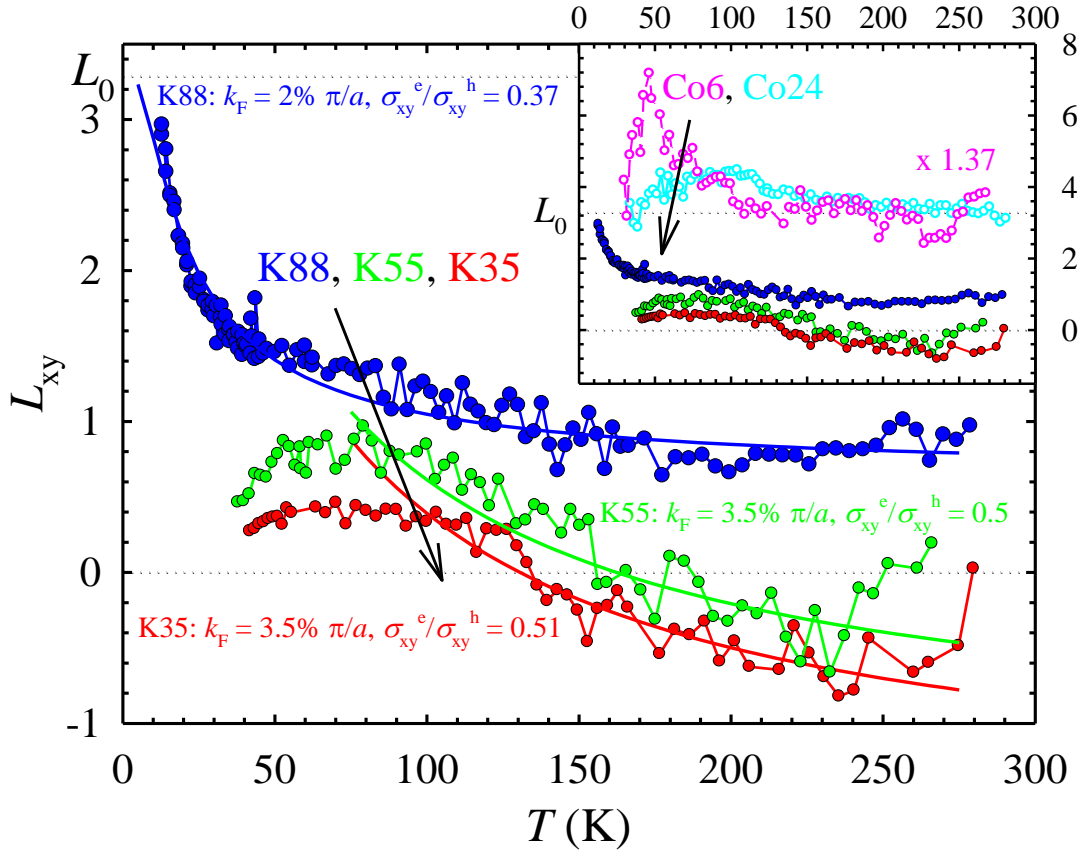

**Figure 5.**

(Color online) The temperature dependences of the Hall Lorenz number for the $Ba_{1-x}K_xFe_2As_2$ series (solid points). Fits to the data with the model described in the text are denoted with solid lines. Inset presents the same set of data along with $L_{xy}(T)$ for $Ba(Fe_{1-y}Co_y)_2As_2$ (hollow points).

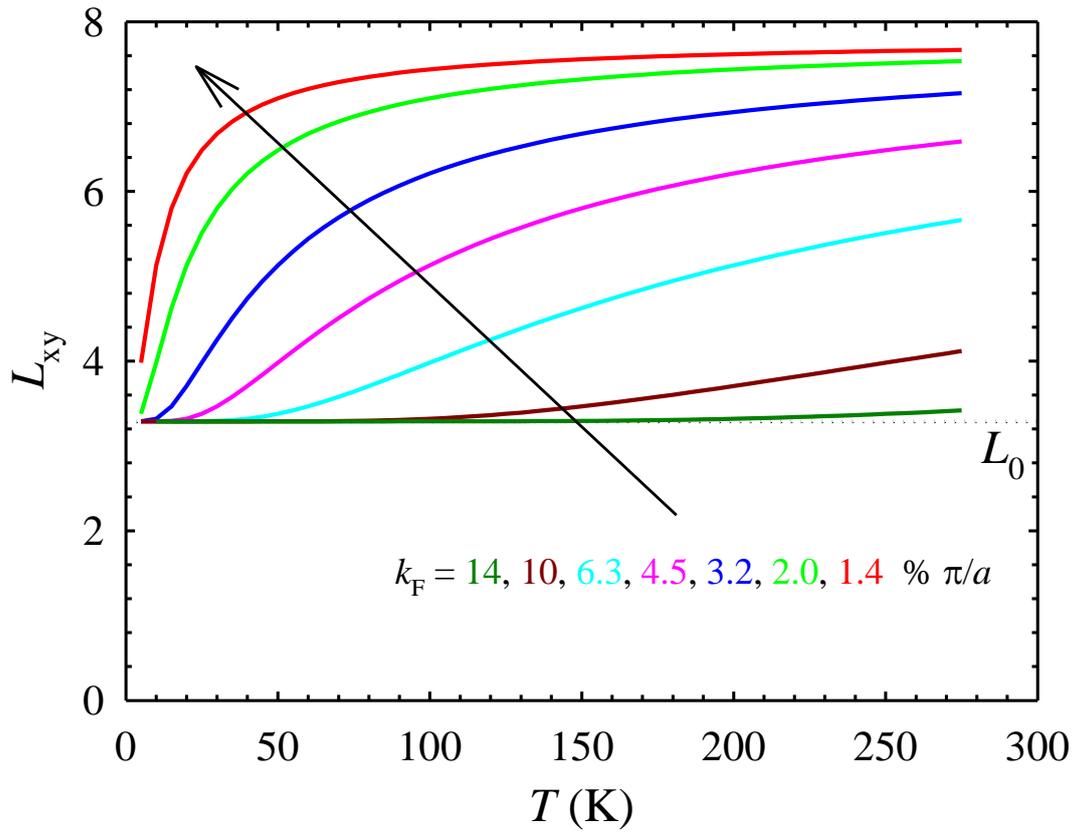

**Figure 6.**

(Color online) The temperature dependences of the Hall Lorenz number calculated for an electron pocket of various size.